\documentclass[runningheads,a4paper]{llncs}

\usepackage{amssymb}
\usepackage{graphicx,url,cite}

\newcommand{\E}{I\!\!E}

\begin{document}

\mainmatter  

\title{Multi-Layer Cyber-Physical Security and Resilience for Smart Grid}


%
%
\author{Quanyan Zhu 
}

\authorrunning{Q. Zhu}
%
\institute{Tandon School of Engineering, New York University, Brooklyn, NY, 11201\\
Email: quanyan.zhu@nyu.edu\\}

%
%

\maketitle

\begin{abstract}
The smart grid is a large-scale complex system that integrates communication technologies with the physical layer operation of the energy systems. Security and resilience mechanisms by design are important to provide guarantee operations for the system. This chapter provides a layered perspective of the smart grid security and discusses game and decision theory as a tool to model the interactions among system components and the interaction between attackers and the system. We discuss game-theoretic applications and challenges in the design of cross-layer robust and resilient controller,  secure network routing protocol at the data communication and networking layers, and the challenges of the information security at the management layer of the grid. The chapter will discuss the future directions of using game-theoretic tools in addressing multi-layer security issues in the smart grid. 
\end{abstract}

\section{Introduction}

The smart grid aims to provide reliable, efficient, secure, and quality energy generation/distribution/consumption using modern information, communications, and electronics technologies. The integration with modern IT technology moves the power grid from an outdated proprietary technology to more common ones such as personal computers, Microsoft Windows, TCP/IP/Ethernet, etc. It  can provide the power grid with the capability of supporting two-way energy and information flow, isolate and restore power outages more quickly, facilitate the integration of renewable energy resources into the grid and empower the consumer with tools for optimizing energy consumption. However, in the meantime, it poses security challenges on power systems as the integration exposes the system to public networks.

Many power grid incidents in the past have been related to software vulnerabilities.
In \cite{example1}, it is reported that hackers have inserted software into the USA power grid, potentially allowing the grid to be disrupted at a later date from a remote location. As reported in \cite{example2}, it is believed that an inappropriate software update has led to  a recent emergency shutdown of a nuclear power plant in Georgia, which lasted for 48 hours. In \cite{example3}, it has been reported that a computer worm, Stuxnet has been spread to target Siemens SCADA systems that are configured to control and monitor specific industrial processes. On November 29, 2010, Iran confirmed that its nuclear program had indeed been damaged by Stuxnet \cite{example4, example5}. The infestation by this worm may   have damaged Iran's nuclear facilities in Natanz and eventually delayed the start-up of Iran's nuclear power plant.

Modern power systems do not have built-in security functionalities, and the security solutions in regular IT systems may not always apply to systems in critical infrastructures. This is because critical infrastructures have different goals and assumptions concerning what needs to be protected, and have specific applications that are not originally designed for a general IT environment. Hence, it is necessary to develop unique security solutions to fill the gap where IT solutions do not apply.

In this chapter, we describe a layered architecture to address the security issues in power grids, which facilitates identifying research problems and challenges at each layer and building models for designing security measures for control systems in critical infrastructures. We also emphasize a cross-layer viewpoint toward security issues in power grids in that each layer can have security dependence on the other layers. We need to understand the tradeoff between the information assurance and the physical layer system performance before designing defense strategies against potential cyber threats and attacks. As examples, we address three security issues of smart grid at different layers, namely, the resilient control design problem at the physical power plant, the data-routing problem at the network and communication layer, and the information security management at the application layers.

The rest of the chapter is organized as follows. In Section 2, we first describe the general multi-layer architecture of cyber-physical systems and the related security issues associated with each layer. In Section 3, we focus on the cyber and physical layers of the smart grid and propose a general cross-layer framework for robust and resilient controller design.  In Section 4, we discuss secure network routing problem at the data communication and networking layers of the smart grid. In addition, we discuss the centralized vs. decentralized routing protocols  and propose a hybrid architecture as a result of the tradeoff between robustness and resilience in the smart grid. In Section 5, we present the challenges of the information security at the management layer of the grid. We conclude finally in Section 6 and discuss future research directions that can follow from the multi-layer model using game-theoretic tools. 

\section{Multi-layer architecture}

Smart grid comprises of physical power systems and cyber information systems. The integration of the physical systems with the cyberspace allows new degrees of automation and human-machine interactions. 
The uncertainties and hostilities existing in the cyber environment have
brought emerging concerns for modern power systems. It is of supreme importance to have a system that maintains state awareness and an acceptable level of operational normalcy in response to disturbances, including threats of an unexpected and malicious nature \cite{Dong1}.  

The physical systems of the power grid can be made to be resilient by incorporating features such as robustness and reliability \cite{Moslehi}, while the cyber components
can be enhanced by many cyber-security measures to ensure dependability, security, and privacy. However, the integration of cyber and physical components does not necessarily ensure overall reliability, robustness, security, and resilience of the power system. The interaction between the two environments can create new challenges in addition to the existing ones.
To address these challenges, we first need to understand the architecture of smart grids. 
The smart grid can be hierarchically organized into six layers, namely, physical layer, control layer, data communication layer, network layer, supervisory layer and management layer. 
The first two layers, physical layer and control layer, can be jointly seen as physical environment of the system. The data communication layer and network layer comprise the cyber environment of the power grid. The supervisory layer together with the management layer constitute the higher-level application layer where services and human-machine interactions take place.

The power plant is at the physical layer, and the communication network and security devices are at the network and communication layers. The controller interacts with the communication layer and the physical layer. An administrator is at the supervisory layer to monitor and control the network and the system. Security management is at the highest layer where security policies are made against potential threats from attackers. SCADA is the fundamental monitoring and control architecture at the control area level. The control center of all major U.S. utilities have implemented a supporting SCADA for processing data and coordinating commands to manage power generation and delivery within the EHV and HV (bulk) portion of their own electric power system \cite{Moslehi2}.

\begin{figure}
\begin{center}
\vspace{-0mm}  \includegraphics[scale=0.33]{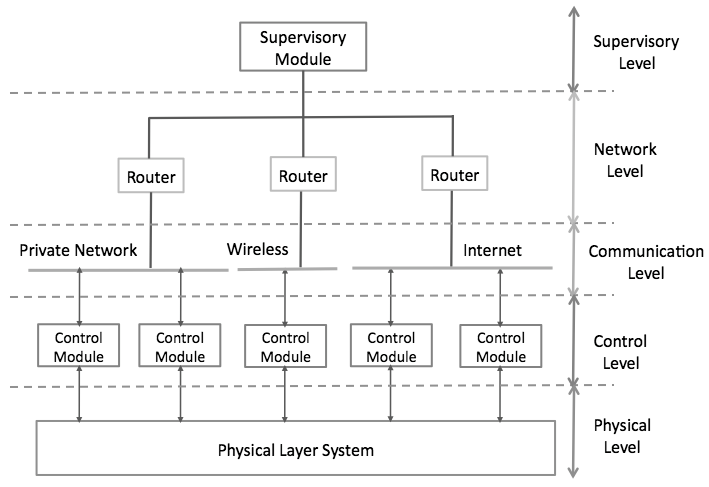}\\
  \caption{A conceptual control system with layering.}\label{LayeringArch}\vspace{-6mm}
\end{center}
\end{figure}

To further describe the functions at each layer, we resort to Figure \ref{LayeringArch}, which conceptually describes a smart grid system with a layering architecture. The lowest level is the physical layer where the physical/chemical processes we need to control or monitor reside. The control layer includes control devices that are encoded with control algorithms that have robust, reliable, secure, and fault-tolerant features. The communication layer passes data between devices and different layers. The network layer includes the data packet routing and topological features of control systems. The supervisory layer offers human-machine interactions and capability of centralized decision-making. The management layer makes economic and high-level operational decisions.

In the following, we identify problems and challenges at each layer and propose problems whose resolution requires a cross-layer viewpoint.

\medskip\noindent{\bf Physical layer}: The physical layer comprises the physical plant to be controlled. It is often described by an ordinary differential equation (ODE) model from physical or chemical laws. It can also be described by difference equations, Markov models, or model-free statistics. We have the following challenges that pertain to the security and reliability of the
physical infrastructure. First, it is important to find appropriate measures to protect the physical infrastructure against vandalism, environmental change, unexpected events, etc. Such measures often need a cost-and-benefit analysis involving the value assessment of a particular infrastructure. Second, it is also essential for engineers to  build the physical systems with more dependable components and more reliable architecture.  It brings the concern on the physical maintenance of the control system infrastructures that demands cross-layer decision-making between the management and physical layers \cite{ZhuCPSWeek1}.
 

\medskip\noindent{\bf Control layer}: The control layer consists of multiple control components, including observers/sensors, intrusion-detection
systems (IDSs), actuators and other intelligent control components. An observer has the sensing capability that collects data from the physical layer and may
    estimate the physical state of the current system. Sensors may need to have redundancies to
    ensure correct reading of the states. The sensor data can be fused locally or sent to the
    supervisor level for global fusion. A reliable architecture of sensor data fusion will be a critical concern.  
An IDS protects the physical layer as well as the communication layer by performing
    anomaly-based or signature-based intrusion detection. An anomaly-based ID is more common for
    physical layer  whereas a signature-based ID  is more common for the packets or traffic at
    the communication layer. If an intrusion or an anomaly occurs, an IDS
    raises an alert to the supervisor or work hand-in-hand with built-in intrusion prevention systems
    (related to emergency responses, e.g., control reconfiguration) to take immediate action.
    There lies a fundamental a trade-off between local decisions versus a centralized decision when intrusions are detected. A local decision, for example, made by a prevention system, can react in time to unanticipated events; however, it may incur a high packet drop rate if the local decision suffers high false negative rates due to incomplete information.
Hence, it is an important architectural concern on whether the diagnosis and control modules need to operate locally with IDS or globally with a
  supervisor. 

\medskip\noindent{\bf  Communication layer}: Communication layer is where we have a communication channel between control layer components or network-layer routers. The communication channel can take multiple forms: wireless, physical cable,
blue-tooth, etc. The communication layer handles the data communication between devices or layers.
It is an important vehicle that runs between different layers and devices. It can often be vulnerable to
attacks such as jamming and eavesdropping. There are also privacy concerns of the data at this layer.
Such problems have been studied within the context of wireless communication networks \cite{zhu1}. However, the goal of critical infrastructure may distinguish themselves from the conventional studies of these issues.

\medskip\noindent{\bf Network layer}: The  network layer concerns the topology of
the architecture. It comprises of two major components: one is network formation, and the other one is routing. We can randomize the routes to disguise or confuse the attackers to achieve certain security or secrecy or minimum delay. Moreover, once a route is chosen, how much data should be sent on that route has long been a concern for researchers in communications \cite{zhu2, ZhuCPSWeek2, ZhuGlobecom2010}. In control systems, many specifics of the data form and rates may allow us to reconsider this problem in a control domain.

\medskip\noindent{\bf Supervisory layer}:
The supervisory layer coordinates all layers by designing and sending appropriate commands. It can be viewed as the brain of the system. Its main function is to perform critical data analysis or fusion to provide an immediate and precise assessment of the situation. It is also a holistic policy maker that distributes resources in an efficient way. The resources include communication resources, maintenance budget as well as control efforts. In centralized control, we have one supervisory module that collects and stores all historical data and serves as a powerful data-fusion and signal-processing center \cite{ZhuDGA2010, ZhuACC2009b}. One key challenge at this layer is to defend against advanced persistent threats which behave stealthily, leverage social engineering, and exploit the vulnerabilities of the computer networks to obtain unauthorized credentials to access the control system networks \cite{stefanGameSec,stefanIEEEAccess}. Hence it is critical to implement security mechanism at this layer to detect intrusive, stealthy and deceptive behaviors, and ensure the integrity of information processing and the availability of critical services.

\medskip\noindent{\bf Management layer}
The management layer is a higher-level decision-making engine, where the decision-makers take an economic perspective toward the resource allocation problems in control systems. At this layer, we deal with
problems such as:
(i) how to budget resources to different systems to accomplish a goal \cite{GordonLoeb2002};
(ii) how to develop policies to maintain data security and privacy \cite{ZhangPrivacy2017,ZhangPrivacy2018};
(iii) how to manage patches for control systems, e.g., disclosure of vulnerabilities to vendors,
    development and release of patches \cite{ZhuCPSWeek3}; and
 (iv) how to invest in cyber insurance to mitigate the losses incurred by the cyberattacks \cite{Zhang2017,Hayel2017}.

Addressing the security challenges at the multiple layers of the smart grid requires a holistic and integrable framework that can capture different system features of the multi-layer cyber-physical security problems. Game theory is a versatile quantitative tool which can be used to model different types of adversarial interactions between a defender and an attacker. For example, at the physical and the control layers, game-theoretic methods can be used to used to design robust and resilient controllers for dynamical systems in an uncertain or adversarial environment \cite{ging,xu3D,xuTrain,xuCDC}. At the supervisory layer, game-theoretic methods can be used to understand spear-phishing attacks \cite{jeff}, insider threats \cite{insider}, and the advanced persistent threats \cite{stefanGameSec,jeffAPT,xuIFAC,Chen2017PES}. At the management layer, game theory serves a primary tool to design strategies for security investment and information disclosure policies.   At the network layer, game theory has been used as a quantitative method for analyzing network security policies and designing defense mechanisms \cite{ZhuACC2011,zhu1,ZhuACC2009a,xu3D}. The wide range of application domains of game theory has made it an ideal tool for developing a unifying framework for a holistic and fundamental understanding of cyber-physical security across different layers of functionalities. In the following section, we will discuss the applications of game-theoretic methods for addressing control, network and management layer problems.

\section{Robust and Resilient Control}

\begin{figure}
\centering
\vspace{-2mm}\includegraphics[scale=0.35]{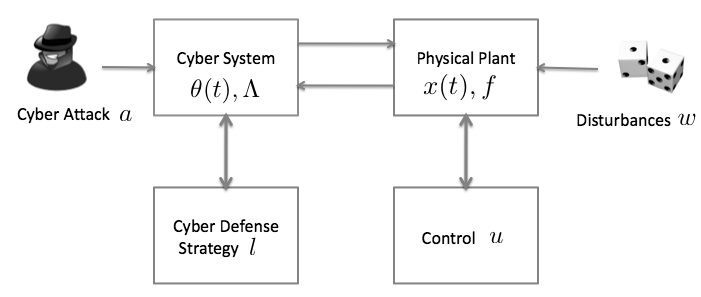}
\vspace{-2mm}\caption{The interactions between the cyber and physical systems are captured by their dynamics governed by the transition law $\Lambda$ and the dynamical system $f$. The physical system state $x(t)$ is controlled by $u$ with the presence of disturbances and noises. The cyber state $\theta(t)$ is controlled by the defense mechanism $l$ used by the network administrator as well as the attacker's action $a$.}\label{SystemIntro}\vspace{-4mm}
\end{figure}

The layered architecture in Figure \ref{LayeringArch} can facilitate the understanding of the cross-layer interactions between the physical world and the cyber world. In this section, we aim to establish a framework for designing a resilient controller for the physical power systems.   In Figure \ref{SystemIntro}, we describe a hybrid system model that interconnects the cyber and physical environments. We use $x(t)$ and $\theta(t)$ to denote the continuous physical state and the discrete cyber state of the system, which are governed by the laws $f$ and $\Lambda$, respectively. The physical state $x(t)$ is subject to disturbances $w$ and can be controlled by $u$. The cyber state $\theta(t)$ is controlled by the defense mechanism $l$ used by the network administrator as well as the attacker's action $a$.  

We view resilient control as a cross-layer control design, which takes into account the known range of unknown deterministic uncertainties at each state as well as the random unanticipated events that trigger the transition from one system state to another. Hence, it has the property of disturbance attenuation or rejection to physical uncertainties as well as damage mitigation or resilience to sudden cyber attacks. It would be possible to derive resilient control for the closed-loop perfect-state measurement information structure in a general setting with the transition law depending on the control action, which can further be simplified to the special case of the linear quadratic problem.

The framework depicted in Figure \ref{SystemIntro} can be used to describe the voltage regulation problem of a power generator subject to sudden faults or attacks. A power system has multiple generators interconnected through a large dynamic network. 
 There are three types of attacks that can be considered: 
\begin{itemize}
\item Sensor Attacks: Attackers can launch a man-in-the-middle attack to introduce a bias to the measured parameters or multiply the sensed value by a constant. 
\item Actuator Attacks:  Attackers can intrude the power control system and disrupt the physical control loops. The attack can cause an error on the generators' output torque, and consequently system dynamics are modified.
\item Controller Attacks: An attacker  can change the control signal sent through  the SCADA system to an extent without being noticed by the system administrator. Consequently, the output of the controller is modified.
\end{itemize}

The framework can incorporate networked control system models to capture different aspects of network effects, for example, sampled-data systems, systems with delayed measurements, and model predictive control systems. The optimal design of the cyber and physical system can be made jointly by viewing each design process as a game-theoretic problem. For example, a zero-sum differential game problem can be used to design a robust controller while a stochastic game model can be used to design a defense strategy. With the joint game design, the framework yields control and defense strategies depend on both cyber states and physical states, and there is the need for the development of advanced computational tools to compute such joint control and defense strategies. 
 
\section{Secure network routing}

One of the challenging issues at the data communication and networking layers of the smart grid in  Figure \ref{LayeringArch} is
the assurance of secure routing of phasor measurement unit (PMU)\index{phasor measurement unit (PMU)} and smart meter (SM) \index{smart meter (SM)} data
in the open network, which is enabled by the adoption of IP-based network technologies. 
It is forecasted that 276 million smart grid communication nodes will be shipped worldwide during the period from 2010 to 2020, with annual shipments increasing dramatically from 15 million in 2009 to 55 million by 2020 \cite{b1}. The current dedicated network or leased-line communication methods are not cost-effective to connect large numbers of PMUs and SMs. Thus, it is foreseen that IP-based network technologies are widely adopted since they enable data to be exchanged in a routable fashion over an open network, such as the Internet \cite{b2, b3,b4,b5,b6}. This will bring benefits such as efficiency and reliability, and risks of cyber attacks as well.  Without a doubt, smart grid applications based on PMUs and SMs will change the current fundamental architecture of communication network of the power grid, and bring new requirements for communication security. Delay, incompleteness, and loss of PMU and SM data will adversely impact smart grid operation in terms of efficiency and reliability. Therefore, it is important to guarantee integrity and availability of those PMUs and SMs data. To meet the QoS requirements in terms of delay, bandwidth, and packet loss rate, QoS-based routing technologies have been studied in both academia and the telecommunications industry \cite{b7, b8,b9,b10}. Unlike video and voice, data communications of PMUs and SMs have different meanings of real-time and security, especially in terms of timely availability \cite{b2,b11,b12,b13,b14,b15}. Therefore, QoS-based and security-based routing schemes for smart grid communications should be studied and developed to meet smart grid application requirements in terms of delay, bandwidth, packet loss, and data integrity.

We can leverage the hierarchical structure of power grids and investigate a routing protocol that maximizes the QoS along the
routing path to the control room. In addition, the data communication rates between the super data concentrator can be optimized
at the penultimate level with the control center. A hybrid structure of routing architecture is also highly desirable to enable the resilience,
robustness, and efficiency of the smart grid.

\begin{figure}
\begin{center}
\vspace{-2mm}  \includegraphics[scale=0.33]{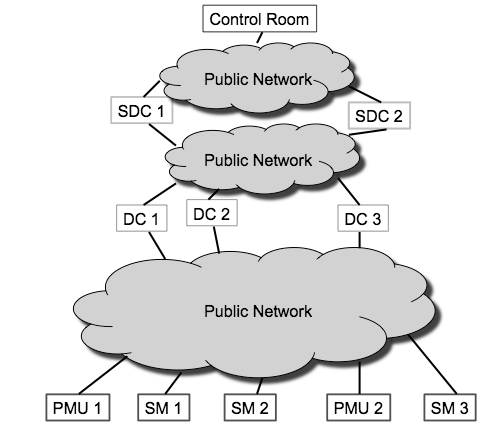}\\
\vspace{-2mm}  \caption{An example of the physical structure of the multi-layer smart grid communication network.}\label{SecureRouting}\vspace{-5mm}
\end{center}
\end{figure}

\medskip\noindent{\bf Hierarchical routing}: 
The smart grid has a multi-layer structure that is built upon the current hierarchical power grid architecture. The end-users, such as households, communicate their power usage and pricing data with a local area substation which collects and processes data from SMs and PMUs. In the smart grid, the path for the measurement data may not be predetermined. The data can be relayed from smaller scale data concentrators (DCs) to some super data concentrators (SDCs) and then to the control room. With the widely adopted IP-based network technologies, the communications between households and DCs can be in a multi-hop fashion through routers and relay devices. The goal of each household is to find a path with minimum delay and maximum security to reach DCs and then substations. This optimal decision can be enabled by the automated energy management systems built in SMs. Figure \ref{SecureRouting} illustrates the physical structure of the smart grid communication network. The PMUs and SMs send data to DCs through a public network. DCs process the collected data and send the processed data to SDCs through (possibly) another public network. 

In the depicted smart grid, the data from a PMU or an SM has to make several hops to reach the control room. The decision for a meter to choose a router depends on the communication delay, security enhancement level, and packet loss rate. In addition, the decisions for a DC to choose an SDC also depends on the same criteria. The communication security at a node is measured by the number of security devices such as firewalls, intrusion-detection systems (IDSs)\index{intrusion-detection system (IDS)} and intrusion-prevention systems (IPSs)\index{intrusion-prevention system (IPS)} deployed to reinforce the security level at that node. We can assign a higher utility to network routers and DCs that are protected by a larger number of firewalls, IDSs/IPSs and dedicated private networks in contrast to public networks. This relatively simple metric only considers one aspect of the control system cyber security. It can be further extended to include more security aspects by considering the authorization mechanisms, the number of exploitable vulnerabilities, potential damages as well as recovery time after successful attacks. The readers can refer to \cite{DHS1,INL, Dong1, Sandia} for more comprehensive metrics.
 
A trade-off with higher security is the latency and packet loss rate incurred in data transmission. A secure network inevitably incurs delays in terms of processing (encrypting/decrypting) and examining data packets. We can model the process of security inspection by a tandem queueing network. Since the arriving packets are inspected by IDS using signature-based or anomaly-based methods to detect malicious behaviors,  each security device can be modeled with a queueing model.  One simple example is the M/M/1 queue whose external arrival rate follows a Poisson process and the service time follows an exponential distribution. The latency caused by the security devices such as IDSs/IPSs is due to the number of pre-defined attack signatures and patterns to be examined \cite{ZhuCPSWeek1, ZhuCDC2009, ZhuACC2009a}. In addition, devices such as IPSs can also lead to high packet loss due to their false negative rates in the detection. 

Furthermore, a node with a higher level of security may be preferred by many meters or routers, eventually leading to a high volume of received data and hence higher level of congestion delay.  Hence it leads to a distributed decision-making problem in which each device determines its route by assessing the tradeoffs between security risks, the congestion delay and the quality of services. This problem can be analyzed using a game-theoretical approach to yield distributed routing decisions in the smart grid \cite{ZhuCPSWeek2,QoS}. The solution concept of mixed Nash equilibrium \cite{basar} as a solution outcome is desirable for two reasons. First, in theory, mixed Nash equilibrium always exists for a finite matrix game \cite{basar} and many learning algorithms such as fictitious play and replicator dynamics can lead to mixed Nash equilibrium \cite{learning, ZhuACC2011, ZhuCDC2011}. Second, the randomness in the choice of routes makes it harder for an attacker to map out the routes in the smart grid.

\medskip\noindent{\bf Centralized vs. decentralized architectures}:
%
%
A centralized routing architecture ensures the global efficiency, and it is robust to small disturbances from SMs and individual DCs or SDCs. However, it is costly to implement centralized planning on a daily basis for a large-scale smart grid.  In addition, global solutions can be less resilient to unexpected failures and attacks as they are less nimble for changes in routes and it takes time for the centralized planner to respond in a timely manner. 

On the other hand, decentralized decision-making can be more computationally friendly based on local information and hence the response time to the emergency is relatively fast. The entire system becomes more resilient to local faults and failures, thanks to the independence of the players and the reduced overhead on the response to unanticipated uncertainties.  However, the decentralized solution can suffer from high loss due to inefficiency \cite{ZhuDGA2010, ZhuACC2009b}. Hence, we need to assess the tradeoff between efficiency, reliability, and resilience for designing the communication protocol between the control stations and the SDCs \cite{Chen2017,Chen2018}.

\section{Management of information security}

The use of technologies with known vulnerabilities exposes power systems to potential exploits. In this section, we discuss information security management which is a crucial issue for power systems at the management layer in Figure \ref{LayeringArch}. The timing between the discovery of new vulnerabilities and their patch availabilities is crucial for the assessment of the security risk exposure of software users \cite{Frei, McQueen}. The security focus in power systems is different from the one in computer or communication networks. The application of patches for control systems needs to take into account the system functionality, avoiding the loss of service due to unexpected interruptions. The disclosure of software vulnerabilities for control systems is also a critical responsibility. Disclosure policy indirectly affects the speed and quality of the patch development. Government agencies such as  CERT/CC (Computer Emergency Response Team/Coordination Center)  currently act as a third party in the public interest to set an optimal disclosure policy to influence the behavior of vendors \cite{CERT}.

\begin{figure*}
\centering
\vspace{-2mm}\includegraphics[scale=0.30]{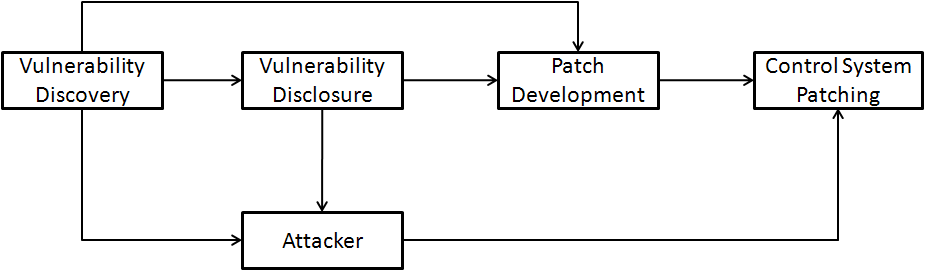}
\vspace{-2mm}\caption{A holistic viewpoint toward vulnerability discovery, disclosure, development, and patching. An attacker can discover a vulnerability or learn it from a disclosure process, eventually influencing the speed of patch application. A discoverer can choose to fully disclose through a forum or report to the vendor or may provide to an attacker. A vulnerability can be disclosed to a vendor for patch development or leaked to the attacker. }\label{VDWholeSystem}\vspace{-2mm}
\end{figure*}

The decisions involving vulnerability disclosure, patch development, and patching are intricately interdependent. In Figure \ref{VDWholeSystem}, we illustrate the relationship between these decision processes. A control system vulnerability starts with its discovery. It can be discovered by multiple parties, for example, individual users, government agencies, software vendors or attackers and hence can incur different responses. The discoverer may choose not to disclose it to anyone, may choose to fully disclose through a forum such as Bugtraq \cite{bugtraq}, may report to the vendor, or may provide to an attacker. 
Vulnerability disclosure is a decision process that can be initiated by those who have discovered the vulnerability. Patch development starts when the disclosure process reaches the vendor and finally a control system user decides on the application of the patches once they become available. An attacker can launch a successful attack once it acquires the knowledge of vulnerability before a control system patches its corresponding vulnerabilities. The entire process illustrated in Figure \ref{VDWholeSystem} involves many agents or players, for example, system users, software vendors, government agencies, attackers. Their state of knowledge has a direct impact on the state of  vulnerability management.

We can compartmentalize the task of vulnerability management into different submodules: discovery, disclosure, development, and patching. The last two submodules are relatively convenient to deal with since the agents involved in the decision making are very specific to the process. The models for discovery and disclosure can be more intricate in that these processes can be performed by many agents and hence specific models should be used for different agents to capture their incentives, utility, resources, and budgets. In \cite{ZhuCPSWeek3}, a dynamic model for control system patching is established to assist users in making optimal patching decisions. It has been shown that the optimal patching intervals are much shorter when risks of potential attacks are taken into account in the system. A dynamic game problem can thus be formulated to study the optimal frequency of patching to minimize the risk of an unpatched control system while an attacker aims to determine the time to launch the attack.

\section{Discussions and Challenges} 
Security issues that arise in the smart grid constitute a pivotal
concern in modern power-system infrastructures. In this chapter, we have discussed a six-layer security architecture
for the smart grid, motivated by the OSI  for the Internet and
PRM models for enterprise and control systems. We have identified the security challenges present
at each layer and pinpointed a holistic viewpoint for security
solutions in the smart grid. The layered architecture facilitates the understanding of the tradeoff between the information assurance at the cyber-related layers and the physical layer system performance. 

We have presented security issues at three different layers. The resilient control design at the physical system is pivotal for modern power systems. We need a hybrid framework in which the occurrence of unanticipated events is modeled by a stochastic switching, and deterministic uncertainties are represented by the known range of disturbances.  It is important to develop new methodologies to take the resilience of physical systems into consideration and  enable a cross-layer control design for modern power grids.

At the data communication and network layers, we need to investigate the secure routing problem in the smart grid, which arises from the adoption of IP-based network technologies due to the wide use of PMUs and smart meters. It is important to leverage the multi-layer structure of power grids and discuss a routing protocol that is based on distributed optimization of the quality-of-service along individual routing paths. The hybrid structure of the routing protocol is desirable to incorporate the desirable features of the centralized and decentralized architectures.

The use of information technologies in power
systems poses additional potential threats due to the frequent
disclosure of software vulnerabilities. At the higher level of the information security management layer, we have discussed  a series of policy-making decisions on vulnerability discovery, disclosure, patch development and
patching. We can use a system approach to understand the interdependencies of these decision processes. 

Game-theoretic methods have provided formal approaches to model the adversarial interactions at multiple layers of the cyber-physical energy system. The game model has taken different forms to design resilient control systems, secure routing, and patching mechanisms. Understanding the security of multi-layer energy system requires a holistic model that integrates the game models to provide an integrated framework for designing cross-layer security strategies. Mitigation of one security threat at one layer can be sometimes more effective than achieving it at other layers. In addition, the success of mitigation of certain attacks can rely on the strategies implemented at other layers. Hence cross-layer game-theoretic models are essential for developing an effective defense under budget constraints for the multi-layer system as a whole. 

More challenges as a result of the multi-layer architecture of the smart grid security involve the integration of game theory, machine learning, control theory, and data-driven approaches for detection, automation, and reconfiguration in the smart grid. In addition to the security problems illustrated in the chapter, there are other security and privacy issues existing at each layer, for example, the jamming and eavesdropping problems at the data communication layer, the user data privacy problem at the management layer, and the system reliability problem at the network layer. Furthermore, the multi-layer framework can be extended to study multi-agent systems. The interactions between sub-systems in the smart grid can reside at the network, communication, and physical layers. It will be interesting to investigate the competition and cooperation for resources at multiple layers.

\end{document}